\documentclass[reprint,amssymb,amsmath, nofootinbib,prl]{revtex4-1}
\usepackage{url}
\usepackage{verbatim}
\RequirePackage{color}
\usepackage{amsmath,amssymb}
\usepackage{epsfig} 
\usepackage{graphicx} 
\usepackage{subfigure}
\usepackage[english]{babel}
\usepackage[colorlinks]{hyperref}
\usepackage{subfigure}
\begin{document}
\title{Application of the confluent Heun functions for finding the quasinormal modes of nonrotating black holes}
\author{Plamen Fiziev\thanks{Department of Theoretical Physics, Sofia University
"St. Kliment Ohridski",  5 James Bourchier Blvd., 1164 Sofia, Bulgaria;JINR, Dubna} and Denitsa Staicova\thanks{Department of Theoretical Physics, Sofia University
"St. Kliment Ohridski",  5 James Bourchier Blvd., 1164 Sofia, Bulgaria}}
\affiliation{Department of Theoretical Physics, Sofia University
"St. Kliment Ohridski",  5 James Bourchier Blvd., 1164 Sofia, Bulgaria}

\email{dstaicova@phys.uni-sofia.bg}
\email{fiziev@phys.uni-sofia.bg}

\begin{abstract}
Although finding numerically the quasinormal modes of a nonrotating black hole is a well-studied question, the physics of the problem is often hidden behind complicated numerical procedures aimed at avoiding the direct solution of the spectral system in this case. In this article, we use the exact analytical solutions of the Regge-Wheeler equation and the Teukolsky radial equation, written in terms of confluent Heun functions. In both cases, we obtain the quasinormal modes numerically from spectral condition written in terms of the Heun functions. The frequencies are compared with ones already published by Andersson and other authors. A new method of studying the branch cuts in the solutions is presented -- the epsilon-method. In particular, we prove that the mode $n=8$ is not algebraically special and find its value with more than 6 firm figures of precision for the first time. The stability of that mode is explored using the $\epsilon$ method, and the results show that this new method provides a natural way of studying the behavior of the modes around the branch cut points.
\end{abstract}

\maketitle
\section{Quasi-normal modes of black holes}
The study of quasinormal modes (QNMs) of a black hole (BH) has long history \cite{QNM,QNM0,QNM1,QNM2,QNM21,special1, special3}. The reason behind this interest is that the QNMs offer a direct way of studying the key features of the physics of compact massive objects, without the complications of the full 3D general relativistic simulations. For example, by comparing the theoretically obtained gravitational QNMs with the frequencies of the gravitational waves, one can confirm or refute the nature of the central engines of many astrophysical objects, since those modes differ for the different types of objects -- black holes, superspinars (naked singularities), neutron stars, black hole mimickers  etc. \cite{special31,NB1,NB2,NB3,NB4,spectra}.

To find the QNMs, one needs to solve the second-order linear differential equations describing the linearized perturbations of the metric: the Regge-Wheeler equation (RWE) and the Zerilli equation for the Schwarzschild metric or the Teukolsky radial equation (TRE) for the Kerr metric and to impose the appropriate boundary conditions -- the so-called black hole boundary conditions (waves going simultaneously
 into the horizon and into infinity)\cite{QNM,QNM0}. Additionally, one requires a regularity condition for the
angular part of the solutions. And then, one needs to solve a connected problem with two complex spectral parameters -- the frequency $\omega$ and the separation constant $E$ ($E=l(l+1)$ -- real for a nonrotating BH, with $l$ the angular momentum of the perturbation). This system was first solved by Chandrasekhar \& Detweiler\cite{QNM} and Teukolsky \& Press \cite{teukolsky} and later developed through the method of continued fractions by Leaver \cite{Leaver}. For more recent results, see also \cite{special3,special1,QNM1,QNM21}.

Because of the complexity of the differential equations, until now, those equations were solved either approximately or numerically meeting an essential difficulty \cite{QNM}. The indirect approaches like the continued fractions method have some limitations and are not directly related with the physics of the problem. The RWE, the Zerilli equation and TRE, however, can be solved analytically in terms of confluent Heun functions, as done for the first time in \cite{Fiziev1,Fiziev3, Fiziev4, Fiziev2}. Imposing the boundary conditions on those solutions {\em directly} (see \cite{Fiziev1,spectra}) one obtains a system of spectral equations \eqref{AE} and \eqref{RE} featuring the confluent Heun functions which can be solved numerically. 

In this article, for the first time we present finding $l$ and $\omega$ {\em directly} in the case for gravitational perturbation $s=-2$ in a Schwarzschild metric, i.e. we solve the RWE and TRE analytically in terms of confluent Heun functions and we use a newly developed method (the two-dimensional generalization of the M\"uller method described in the internal technical report \cite{arxiv}) to solve the system of two transcendental equations with two complex variables. Then we use the epsilon method to study the stability of the solutions with respect to small variations in the phase condition.  

The results are compared with already-published ones and are found to coincide with at least 8 digits for the RWE and 6 digits for the TRE. For the first time, the so-called algebraically special mode $n=8$ is evaluated with precision of more than 6 digits, and it is shown to have a nonzero real part. This firmly refutes the hypothetical relation of this mode with the algebraically special once. Also demonstrated is the nontrivial dependence on $\epsilon$ of the first 11 modes in both cases.

\section{General form of the equations}
The angular equation for both cases is the solution of the Teukolsky angular equation when there is no rotation ($a=0$):
\begin{align}
S(\theta)&\!=\!(\cos(\theta) \!-\! 1) (\cos(\theta)\! +\! 1) \text{LegendreP}(l, 2, \cos(\theta))\!=\!0
\label{AE}
\end{align}
\noindent where $\theta \in [0,\pi]$ is the angle. The results for the QNMs should be independent
of the choice of $\theta$ in the spectral conditions. In our numerical experiments, we use $\theta=\pi-10^{-7}$. 

The general form of the radial equations is obtained from the solutions of the RWE and TRE written in terms of the confluent Heun functions according to \cite{Fiziev3}, on which the black hole boundary conditions have been imposed. The choice of the local solution in terms of the Heun function takes into account the boundary condition on the horizon. Then, it remains to impose the following boundary condition on the space infinity (for details see \cite{spectra,Fiziev3}):
\begin{align}
&R=r_\infty^{p} \text{HeunC}(\alpha,\beta,\gamma,\delta,\eta,1-r_\infty)=0,
\label{RE}
\end{align}
\noindent where $\text{HeunC}$ is the confluent Heun function as defined in \textsc{maple} and the parameters $\alpha,\beta,\gamma,\delta,\eta \, \,\text{and}\,\, p$ differ for the two equations. The values of the parameters when the BH mass is $M=1/2$ and, if we choose $|r_\infty|=20$ which turns out to be large enough to simulate numerically the actual infinity,  are (\cite{Fiziev3,Fiziev2}):

\begin{enumerate}
 \item for the solutions of the Regge-Wheeler Equation:
\begin{align*}
&\alpha=-2\,i\omega, \beta=2\,i\omega, \gamma=4,\\
&\delta=-2\,{\omega}^{2},\eta=4-l-{l}^{2}+2\,{\omega}^{2},\\
&r_\infty=20\,{{\rm e}^{-i \left( 1/2\, \left( 1+{\epsilon}
 \right) \pi +{\it arg} \left( \omega \right)\right)}}, p=3,
\end{align*}

\item for the solutions of the Teukolsky Radial Equation:
\begin{align*}
&\alpha=-2\,i\omega, \beta=2+2\,i\omega, \gamma=2,\\
&\delta=-4i\omega-2\,\omega^2,\eta=2\,\omega^2+4i\omega-A,\\
&r_\infty=20\,{{\rm e}^{-i \left( 1/2\, \left( 1+\epsilon
 \right) \pi +{\it arg} \left( \omega \right)\right)}}, p=5,
\end{align*}
\noindent where $A=l(l+1)-s(s+1)$ is the separation constant. The parameters were obtained by solving the Teukolsky radial equation and substituting $a=0,$ and they are clearly different from those in the Regge-Wheeler case. Hence, it is important to check whether both methods give the same results for QNM and with what precision.  

\end{enumerate}
\section{The epsilon method}
For values of the parameters $\alpha,\, \beta,\,\gamma,\,\delta,\,\eta$ of general type, the confluent Heun function $\text{HeunC}(\alpha, \beta,\gamma,\delta,\eta, z)$ has branching points in the complex z-plane at the singular points $z=1$ and $z=\infty$. In the \textsc{maple} package, as a branch cut is chosen the semi-infinite interval $(1,\infty)$ on the real axis. The presence of the branch cut may lead to the disappearance of some modes or their translation, since by changing the phase of the complex variable $r$, we may make a transition to another sheet of the multivalued function. To avoid this, we use the epsilon method with which one can find the correct sheet and remain on it. This is done by introducing a small variation ($\mid\!\epsilon\!\mid<\!1$) in the phase condition $\arg(r)\!+\!\arg(\omega)\!=\!-\!\pi/2$ (defined by the direction of steepest descent, see \cite{Fiziev1}), with which one can move the branch cuts farther from the roots and thus avoid the jump discontinuity in the function.  For more information on the epsilon method and the numerical procedures, see \cite{arxiv}.

\section{Numerical Results}
From the angular equation \eqref{AE}, it is clear that it can be solved explicitly without solving the system  \eqref{AE} and \eqref{RE} and the values of $l$ are known: $l=2,3,\ldots$. In this paper, only the first value, $l=2$, is used to find the QNMs with both radial equations. One can then either solve only the radial equations or  solve the systems \eqref{AE} and \eqref{RE} with the appropriate values of the parameters. If one solves the problem as a two-dimensional system, making calculations with 15 digits of precision (and 32 software floating-point digits), one obtains as expected, $l=1.99(9)+1\times 10^{-17}i$ with the first digit different from digit 9 being the 17th. 

The numerical results for the frequencies are summed in Table \ref{table1}.

\begin{table}[!h]
\footnotesize
\begin{tabular}{|l | l  | l | l|}
 \hline $n$  & $\omega$ from the Regge-Wheeler Eq. &$\omega$ from the Teukolsky Eq. \\ \hline
0&0.7473433688+0.1779246316i&0.7473433676+0.1779246260i\\
1&0.6934219937+0.5478297504i&0.6934219698+0.5478298839i\\
2&0.6021069092+0.9565539668i&0.6021069568+0.9565538786i \\
3&0.5030099245+1.4102964056i&0.5030097036+1.4102966442i\\
4&0.4150291600+1.8936897821i&0.4150291670+1.8936897747i\\
5&0.3385988052+2.3912161094i&0.3385987682+2.3912160831i\\
6&0.2665046794+2.8958212549i&0.2665047149+2.8958212406i\\
7&0.1856446653+3.4076823515i&0.1856446394+3.4076823843i\\
8&-0.0306490371+3.9968237195i&-0.0306490242+3.9968236554i\\
9&0.1265269702+4.6052896060i&0.1265270059+4.6052895329i\\
10&0.15310679658+5.1216534769i&0.1531069231+5.1216532271i\\
\hline
\end{tabular}
\caption{A list of the frequencies obtained for the QNMs of Schwarzschild black hole using the Regge-Wheeler equation and the Teukolsky equation. The modes with $n<5$ are found for $\epsilon=0$, modes from $n\geq 5$ -- with $\epsilon=-0.3$. The first 5 frequencies ($n=0-4$) were obtained also by Fiziev in \cite{Fiziev1} using exact solutions of RWE in the Heun functions}
\label{table1}
\end{table}
From the table, one can see that the frequencies from the two types of equations coincide with at least 6 digits. A comparison between the RWE frequencies and the ones published by Andersson \cite{Q_N_M}, published in \cite{arxiv} shows that the difference between the two results is smaller than $5\times10^{-8}$ in most cases and is due to the numerical reasons. 

There are two important results from this study. First, as seen from Table \ref{table1} for both the RWE and the TRE, the mode number 8 has a small but nonzero real part. According to Leaver's evaluations this mode should be equal to $0+3.998000i$ \cite{Leaver}, with an exactly zero real part, if it is to correspond to the so-called {\em algebraically special mode}.

Algebraically special (AS) modes have a special place in the QNM studies \cite{QNM}. The Andersson method is not applicable for them and these are excluded from his consideration. Berti, Cardoso and Starinets (\cite{special3, special1}) make a review on the results so far concerning these modes. Theoretically the 9th mode ($n=8$) should be purely imaginary with value $4i$, if it indeed corresponds to the AS case. In our results, even though purely imaginary modes do not pose a problem for the method, the real part of the 9th mode is distinctly not zero, and it has at least 7 stable digits when changing $\epsilon$ in the interval discussed below for both RWE and TRE. This clearly shows that this mode does not agree with the hypothesis for the AS mode, which is to be expected since the AS mode should correspond to different boundary conditions -- those of the so-called totally-transmission modes (\cite{special2}. 

The second important result is the dependence of the frequencies $\omega_n$ on $\epsilon$. The direction of steepest descent is supposed to be the optimal direction in which the solutions satisfy the black hole boundary conditions on infinity in the first term approximation for asymptotic series for the Heun functions \cite{Fiziev1}. The validity of steepest descent method in its simplest form for the radial equations \eqref{RE} in both cases under variations in this condition, however, is still an open problem studied here for the first time. 

Using the $\epsilon$ method, one can explore the intervals for $\epsilon$ in which each mode can be found. The results for both RWE and TRE, as expected, coincide. Generally, the intervals into which each mode can be found narrow down when increasing $n$. While for the first 5 modes it is possible to find $\omega_n=\pm|\Re(\omega_n)|+\Im(\omega_n)i$ for positive and negative values of $\epsilon$ in a certain interval \footnote{The ranges where each mode is found depend on $\epsilon$ as follows: for $n=0$, $\epsilon \in [\mp 0.8, \pm 0.75]$, for $ n=1,\, \epsilon \in [\mp 0.8, \pm 0.45]$, for $ n=2,\, \epsilon \in [\mp 0.8, \pm 0.25]$, for $ n=3,\,\epsilon \in [\mp 0.8, \pm 0.1]$, for  $ n=4,\, \epsilon \in [\mp 0.8, 0]$, where the first sign corresponds to frequencies with a positive real part and the second sign to those with negative real parts. The imaginary parts for each mode $n$ coincide.}, for $n>4$, (but $n\neq8$) the modes with a positive real part can be found only for negative values of $\epsilon$, and the dependency becomes $\omega_n(\epsilon)=-\text{sgn}(\epsilon)|\Re(\omega_n)|+\Im(\omega_n)i$. 

For $n=8$, the mode has different behavior with respect to $\epsilon$ -- for  $\epsilon \in [-0.75,-0.1]$, one finds a mode with {\em negative} real part and vice versa: ($\omega_{n=8}\!=\!\text{sgn}(\epsilon)\,0.030649006+3.996823690i$).

The so-found relation $\omega_n(\epsilon)$ needs to be examined further. For the case $n = 8$, similar (to some extent) behavior was mentioned also in \cite{special2,AS} (and discussed in \cite{special1}). It was suggested that there are two AS modes which are symmetrical to the imaginary axis and perhaps may be related with the branch cut in the asymptotic of the RWE potential when $\omega$ is purely imaginary. Using the $\epsilon$ method applied on the asymptotics of the confluent Heun functions, one can directly obtain the place of the branch cut on the real axis as a function of $\epsilon$  and they can be easily visualized plotting the solution $R^{\pm}$. Therefore, the use of the confluent Heun functions and the $\epsilon$ method offers a direct way to examine the solutions and their properties in relation to the branch cut in the complex r-plane, something that cannot be readily done in the continued fractions method generally used to obtain the QNMs. 

\begin{figure}[!h]
 \centering
$\begin{array}{cc}
 \subfigure[]{\includegraphics[height=120px,width=110px]{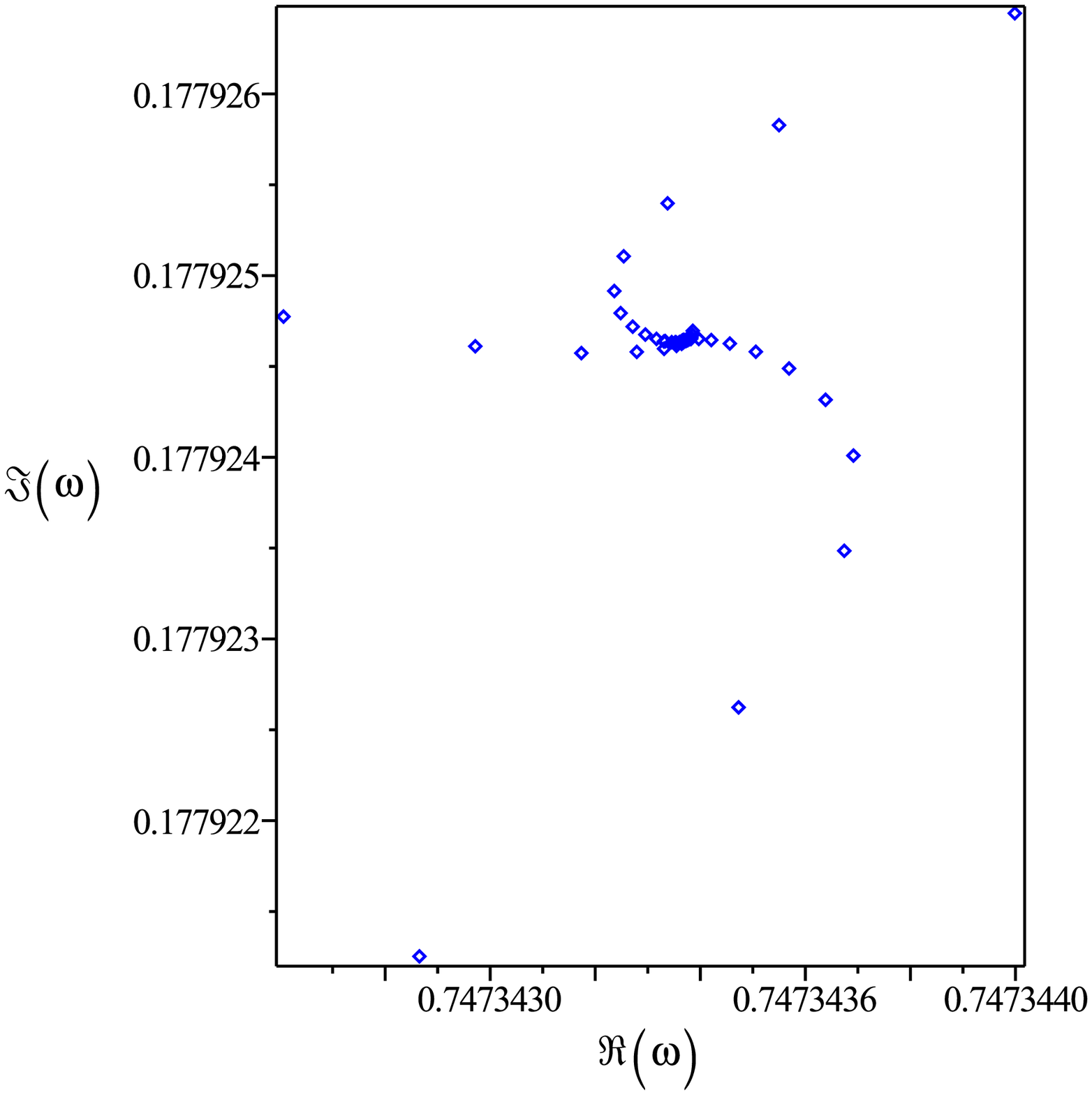}} &
 \subfigure[]{\includegraphics[height=120px,width=110px]{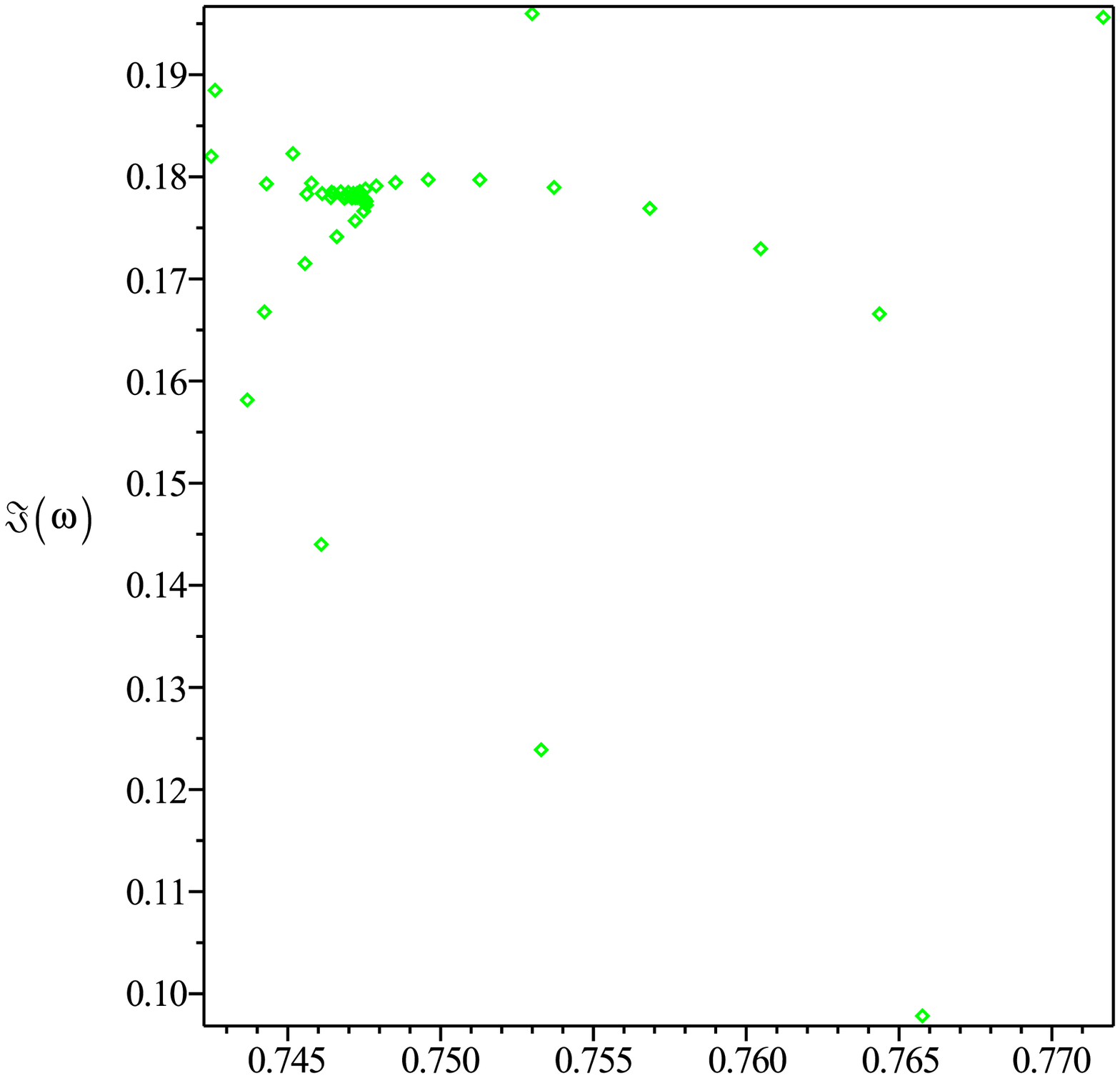}} \\
  \subfigure[]{\includegraphics[height=120px,width=110px]{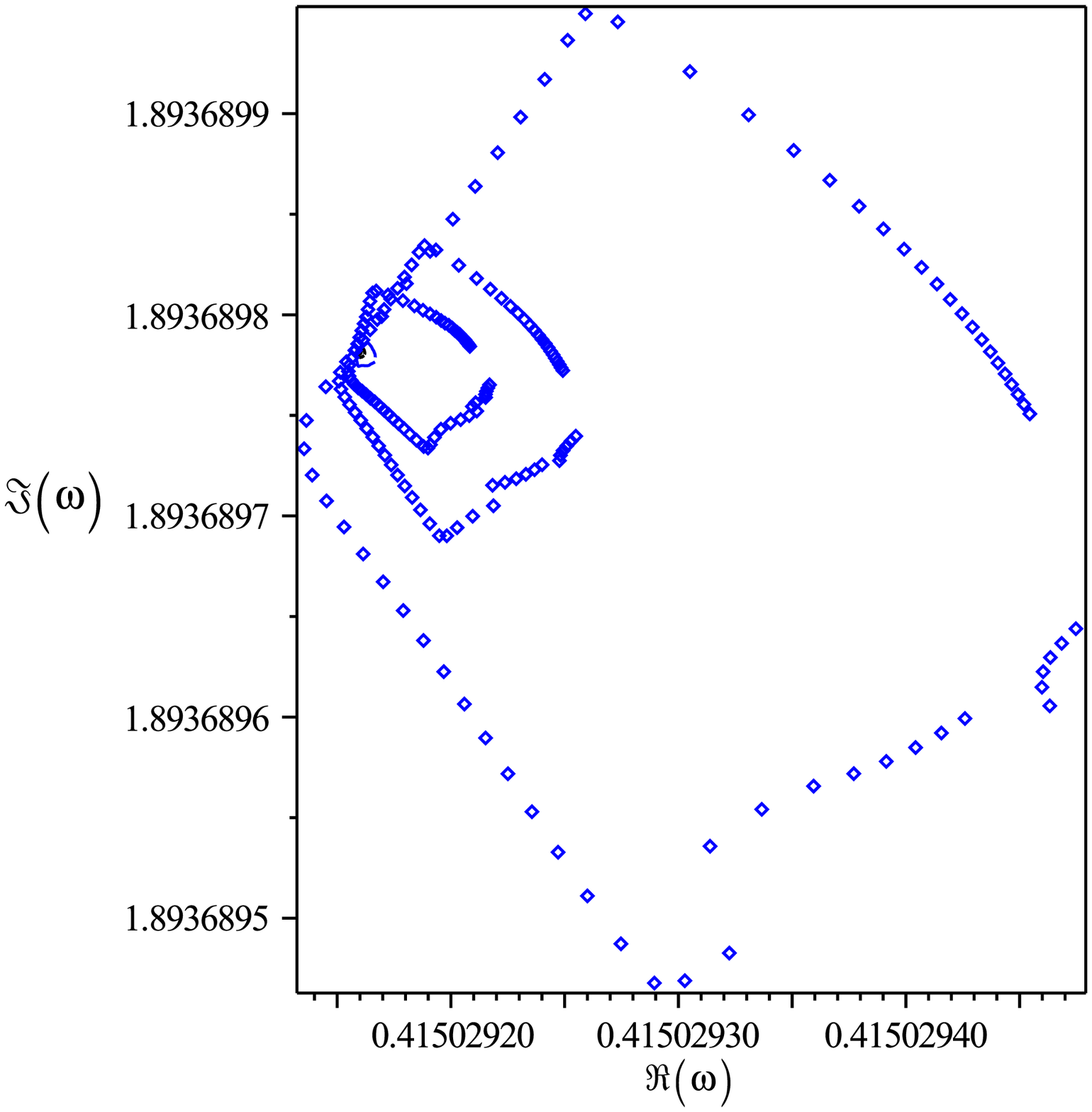}}&
 \subfigure[]{\includegraphics[height=120px,width=110px]{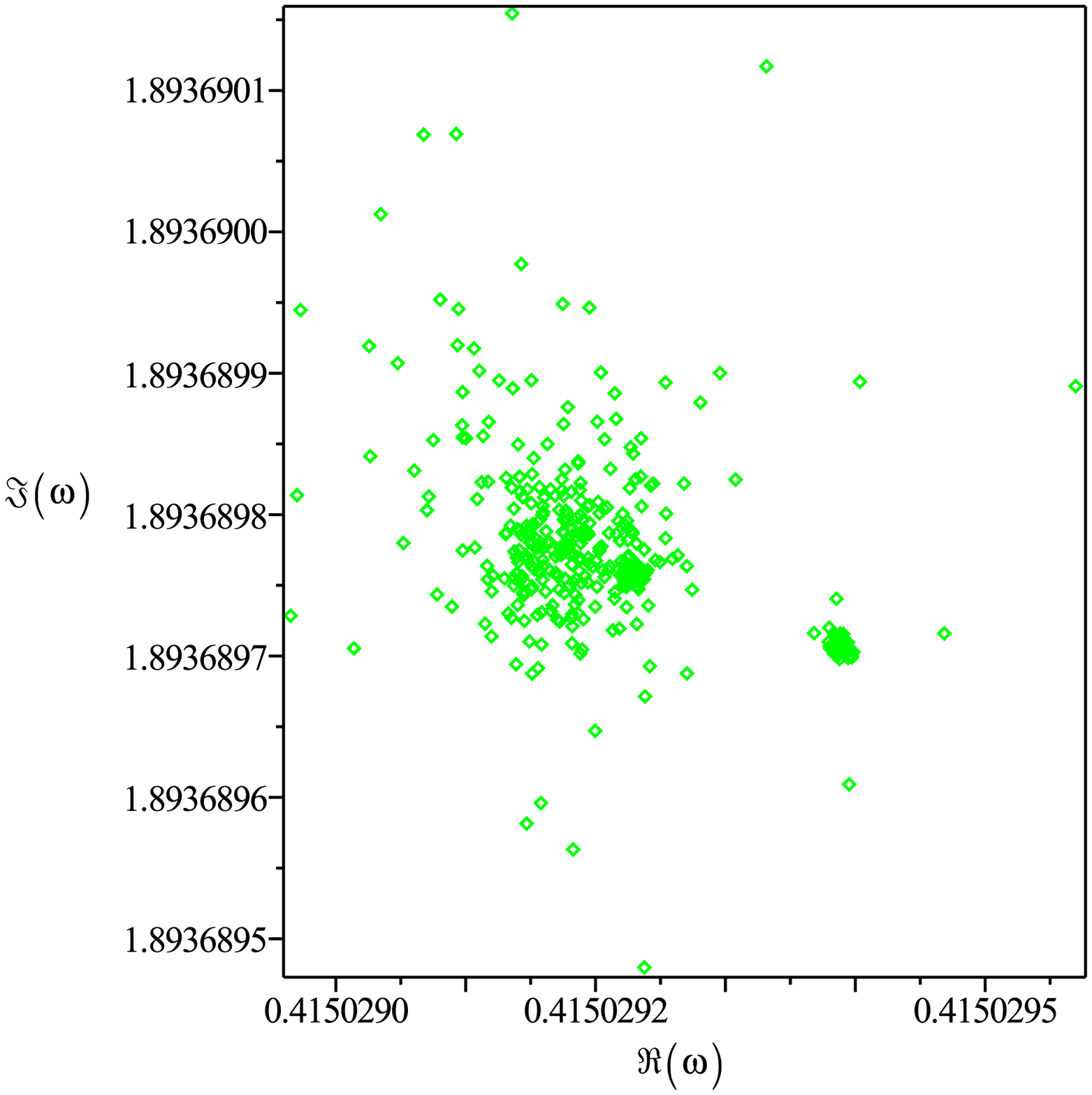}}\\
\end{array} $
\caption{Complex plots of the {\em scaled} QNMs from the two equations in the appropriate intervals for $\epsilon$:
a)RWE $n=0..3$ b)TRE $n=0..3$ c) RWE $n=4..10$ d)TRE $n=4..10$. Clearly, while for $n=0..3$ the QNMs from the two equations give similar results, for $n>4$, the variations in the frequencies from TRE happen on a much smaller scale and appear chaotic.}
 \label{fig1}
 \end{figure}

\begin{figure}[!h]
\centering
 \subfigure[]{\includegraphics[scale=0.15]{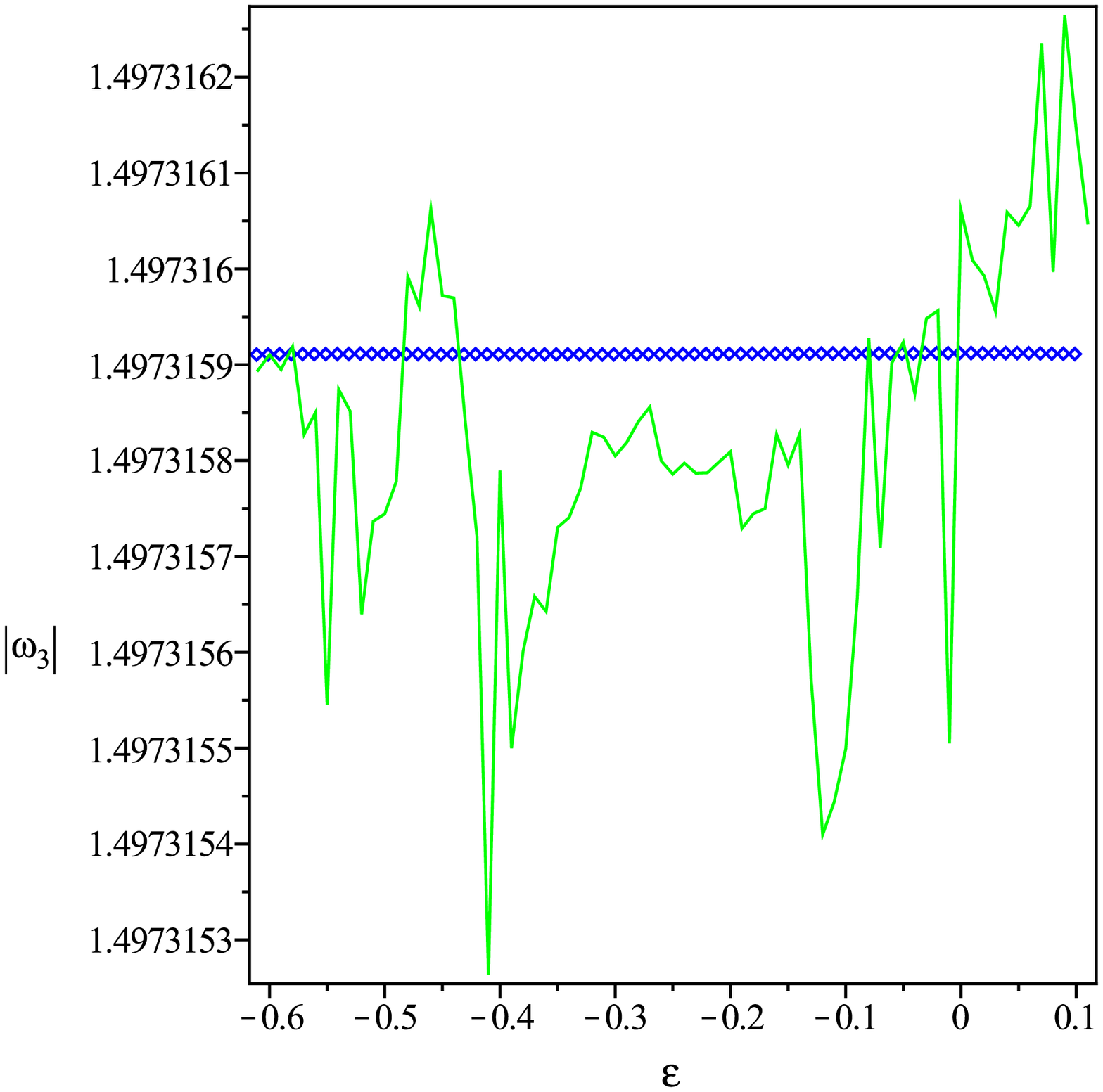}}
 \subfigure[]{\includegraphics[scale=0.15]{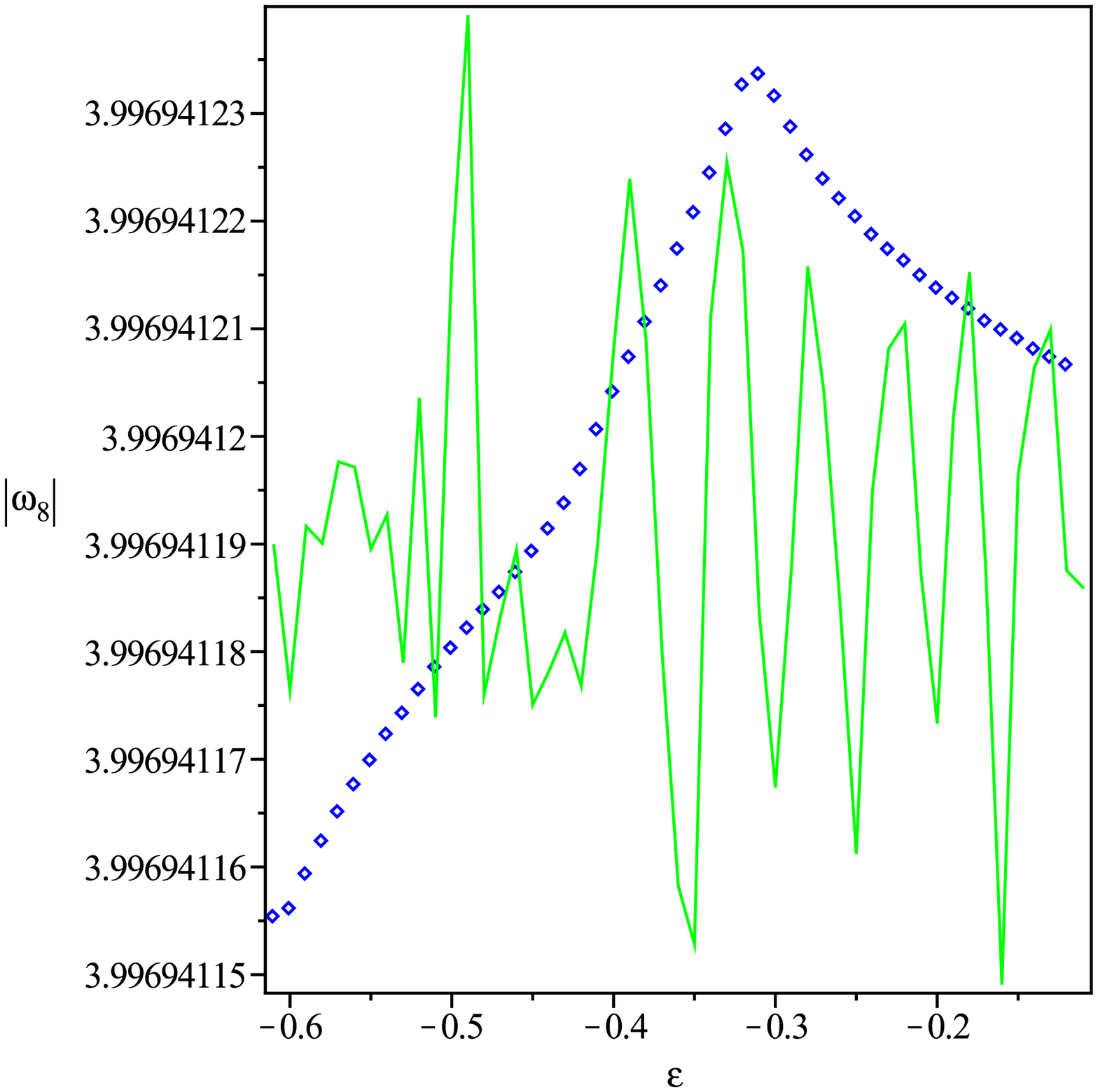}}
 \subfigure[]{\includegraphics[scale=0.15]{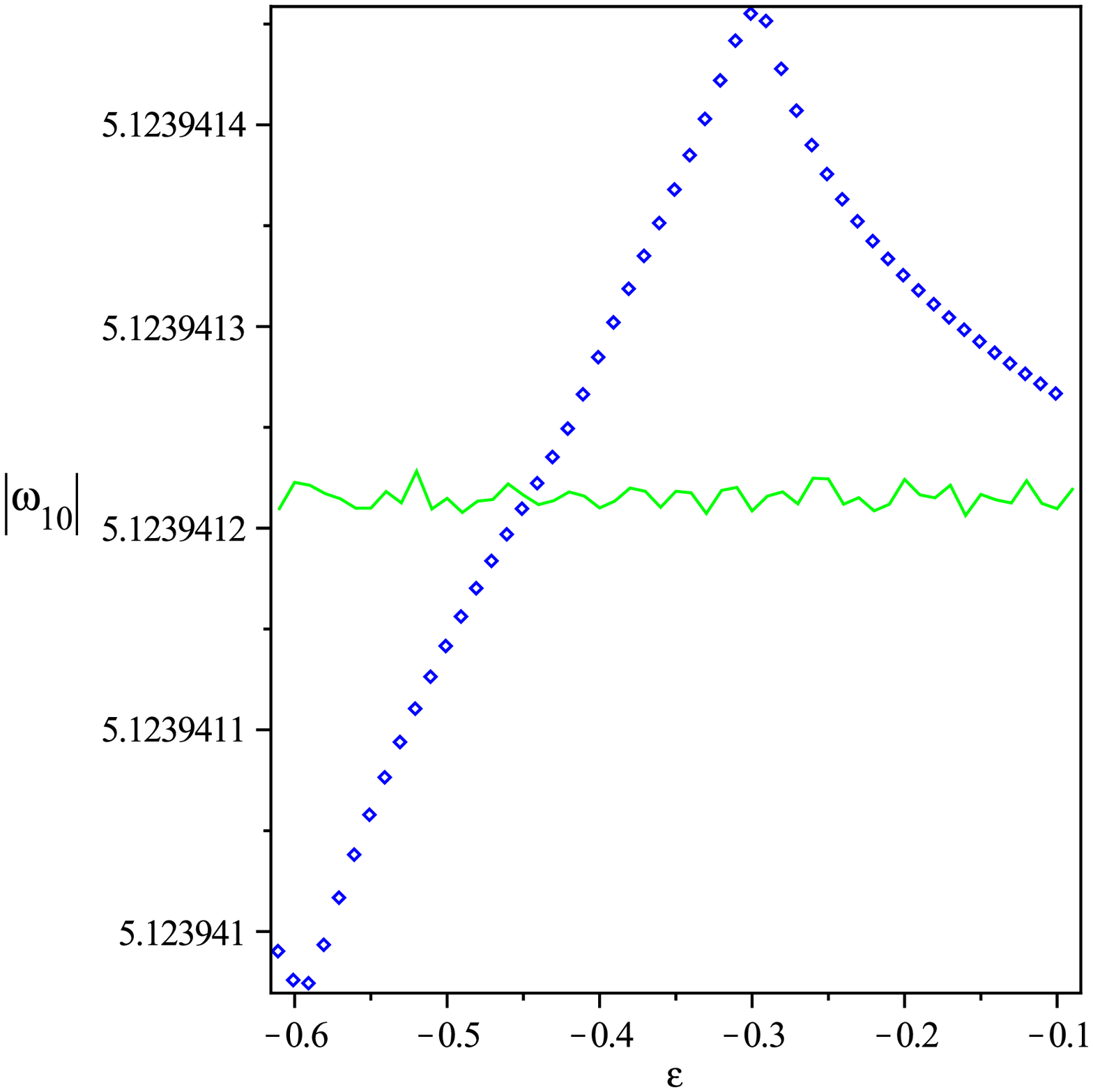}}
\caption{Plots of $abs(\omega)(\epsilon)$, where the solid line corresponds to the TRE modes and the dots -- to the RWE modes for a) n=3, b) n=8, c) n=10. One can see the evolution of the amplitudes and the periods in both cases, when $n$ increases from 3 to 10. }
 \label{fig2}
\end{figure}

Further exploration of the dependence $\Re(\omega_n)(\epsilon)$ (or $\Im(\omega_n)(\epsilon)$) in the intervals mentioned above shows that, for both the RWE and the TRE, it is approximately a periodic function with amplitude $A$ and period $L$ which change with $n$ in a nontrivial way (Fig.\ref{fig1} and Fig. \ref{fig2}). For $n<4$, from the RWE and the TRE one obtains $A_{TRE}\approx 10^{-6}\approx 10^3 \times A_{RWE}$, $L_{RWE}\approx L_{TRE}
\approx 0.4$ and those values remain approximately constant with respect to $n$ ($n<4$). For $n\ge 4$, the dependence of $A$ and $L$ on $n$ becomes more pronounced: the amplitudes and the periods of the RWE increase with $n$ until they reach $A_{RWE}\approx10^{-6}$, $L_{WRE}\approx 0.6$ for $n=10$. For the TRE, the amplitude and the period decrease to $A_{TRE}\approx10^{-8}$, $L_{TRE}\approx0.05$. For $n=8$, the two periodic behaviors have approximately equal amplitudes $\approx 10^{-7}$. Those results hint that, although the so-obtained frequencies are stable with at least 6 digits with respect to $\epsilon$, there is also some finer dependence, the origin of which should be carefully investigated. 

\section{Conclusion}
In this paper, were presented the QNMs for a Schwarzschild BH obtained from the RWE and the TRE, by solving the differential equations analytically in terms of confluent Heun functions. The QNMs from the TRE for the case $s=-2$ were calculated for the first time and were found to coincide with the well-known QNMs from the RWE with precision of 6 digits. 

We demonstrated a new method for studying the stability of the QNM calculations.  The results show nontrivial dependence on small variation in the phase condition (the $\epsilon$ method) which requires additional investigation. 

For the first time, the mode $n=8$ was obtained directly from the spectral condition on the exact analytical solutions of RWE and TRE and was found to have a nonzero real part, which proves that this mode is not the algebraically special mode. The mode in question is stable with 6 digits of significance with respect to changes in $\epsilon$, which proves that its real part is indeed not zero. 

Those results presented here show the strength of using confluent Heun functions to find QNMs of nonrotating BHs and are encouraging in continuing this work in finding QNMs of rotating BHs.

\section{Acknowledgements}
This article was supported by the Foundation "Theoretical and
Computational Physics and Astrophysics", by the Bulgarian National Scientific Fund
under contracts DO-1-872, DO-1-895, DO-02-136, and Sofia University Scientific Fund, contract 185/26.04.2010.

\section{Author Contributions}
P.F. chose the evaluation of the QNMs of non-rotating BHs as a test of the two-dimensional M\"uller algorithm, proposed the epsilon method, as a generalization of the previous work and he supervised the project.

D.S. is responsible for the calculation of QNMs, based on the implementation of confluent Heun functions and for the exploration and analysis of the $\epsilon$-method in the sector of the complex r-plane where QNMs can be found.

Both authors discussed the results at all stages. The manuscript was prepared by D.S. and edited by P.F..

\makeatletter
\let\clear@thebibliography@page=\relax
\makeatother

\end{document}